\documentclass[conference]{IEEEtran}
\usepackage{ifpdf}
\usepackage[nospace]{cite}
\usepackage{graphicx}
\usepackage{color}

\usepackage{bm}
\usepackage[cmex10]{amsmath}
\usepackage{amssymb}
\usepackage{array}
\usepackage{syntonly}
\usepackage{amsthm}
\usepackage{amsfonts}
\usepackage{epsfig}
\usepackage{latexsym}
\newcommand{\Define}{\stackrel{\triangle}{=}}

\DeclareMathOperator{\diagonal}{diag}
\DeclareMathOperator{\Tr}{Tr}

\usepackage{fixltx2e}
\usepackage{url}

\theoremstyle{definition} 
\theoremstyle{definition} 
\theoremstyle{definition} 

\begin{document}
\title{{\LARGE Transmitter Optimization in Slow Fading MISO Wiretap Channel} }
\author{
{\large Sanjay Vishwakarma and A. Chockalingam} \\
       Email: sanvish1975@gmail.com, achockal@ece.iisc.ernet.in \\
      {\normalsize Department of ECE, Indian Institute of Science, Bangalore 560012, India}
}
\maketitle
\begin{abstract}
In this paper, we consider the transmitter optimization problem
in slow fading multiple-input-single-output (MISO) wiretap channel. 
The source transmits a secret message intended for $K$ users in the presence of $J$ 
non-colluding eavesdroppers,
and operates under a total power constraint.
The channels between the source and all users and eavesdroppers are assumed to be slow fading,
and only statistical channel state information (CSI) is known at the source. 
For a given code rate and secrecy rate pair of the wiretap code, 
denoted by $(R_{D}, R_{s})$,
we define the non-outage event as the joint event of the link 
information rates to $K$ users be greater than or equal to $R_{D}$ and the link 
information rates to $J$ eavesdroppers be less than or equal to $(R_{D} - R_{s})$.
We minimize the transmit power subject to the total power constraint
and satisfying the probability of the non-outage event
to be greater than or equal to a desired threshold $(1-\epsilon)$.
\end{abstract}
{\em keywords:}
{\em {\footnotesize
Physical layer security, MISO wiretap channel, secrecy rate, multiple eavesdroppers, slow fading. 
}} 
\IEEEpeerreviewmaketitle

\section{Introduction}
\label{sec1}
With growing applications on wireless networks,
there is a need to provide security, along with reliability, 
from being eavesdropped, which can
easily happen due to the broadcast nature of the wireless transmission.
Wyner, in his work in \cite{ir1}, showed that a message could be transmitted at a rate
called secrecy rate, at which the legitimate user could decode the message reliably
whereas the eavesdropper could be kept entirely ignorant. The wiretap channel model
in \cite{ir1} was physically degraded and discrete memoryless.
Later, the work in \cite{ir1} was extended to more 
general broadcast channel and Gaussian channel in \cite{ir2} and \cite{ir3}, respectively.
Subsequent extension to various multi-antenna wireless wiretap channels 
and the corresponding achievable secrecy rates 
and secrecy capacities have been reported by many authors, e.g., \cite{ir4, ir5, ir6, ir7, ir8, ir9, ir10, ir11}. 

In \cite{ir13}, secrecy capacity of a quasi-static single-antenna 
Rayleigh fading channel in terms of outage probability has been characterized.
Outage probability characterization of the secrecy rate of multiple-input-single-output (MISO) wiretap channel
with artificial noise has been reported in \cite{ir20, ir19, ir18}, and that of
amplify-and-forward relay channel has been reported in \cite{ir14}.
Motivated by the need for outage probability characterization of secrecy rate in MISO wiretap channel,
in this paper, we consider the transmitter optimization problem
in slow fading MISO wiretap channel. 
The source transmits a secret message intended for $K$ users in the presence of $J$ 
{non-colluding} eavesdroppers, and
operates under a total power constraint.
The channels between the source and all users and eavesdroppers are assumed to be slow fading.
Only statistical channel state information (CSI) is assumed to be known at the source. 
{For a given code rate and secrecy rate pair of the wiretap code, 
denoted by $(R_{D}, R_{s})$,
we define the non-outage event as the joint event of the link 
information rates to $K$ users be greater than or equal to $R_{D}$ and the link 
information rates to $J$ eavesdroppers be less than or equal to $(R_{D} - R_{s})$.
We minimize the transmit power subject to the total power constraint
and satisfying the probability of the non-outage event
to be greater than or equal to a desired threshold $(1-\epsilon)$.}
We obtain the achievable ($R_{D}, R_{s}$) region
and the transmit beamforming vector.
{We note that we differ from the reported works in \cite{ir20, ir19}, 
which also consider multiple eavesdroppers scenario, in following aspects:
$i)$ number of users $K$ can be more than one, $ii)$ 
only statistical CSI of the users channels are known, 
and $iii)$ channel covariance matrices of all users and eavesdroppers can be arbitrary positive 
semidefinite matrices.}

$\bf{Notations:}$ $\boldsymbol{A} \in 
\mathbb{C}^{N_{1} \times N_{2}}$ implies that $\boldsymbol{A}$ is a complex 
matrix of dimension $N_{1} \times N_{2}$. 
$\boldsymbol{A} \succeq \boldsymbol{0}$ and $\boldsymbol{A} \succ \boldsymbol{0}$ imply
that $\boldsymbol{A}$ is a positive semidefinite matrix and
positive definite matrix, respectively.
Identity matrix is denoted by $\boldsymbol{I}$.
Transpose and complex conjugate 
transpose operations are denoted by $[.]^{T}$ and $[.]^{\ast}$, respectively.
$\mathbb{E}[.]$ denotes the expectation operator, and
$\parallel \hspace{-1mm}. \hspace{-1mm}\parallel$ denotes the 2-norm operator.
$\diagonal(\boldsymbol{a})$ denotes a diagonal matrix with elements of the vector $\boldsymbol{a} \in 
\mathbb{C}^{N \times 1}$ on its diagonal. { Trace of matrix 
$\boldsymbol{A} \in \mathbb{C}^{N_{} \times N_{}}$ is denoted by $\Tr( \boldsymbol{A} )$.
$\boldsymbol{h}_{} \in \mathbb{C}^{N_{} \times 1} \sim \mathcal{CN}(\boldsymbol{0}, \boldsymbol{H}_{})$ implies that 
$\boldsymbol{h}_{}$ is a circularly symmetric complex Gaussian random vector
with mean vector $\boldsymbol{0}$ and covariance matrix $\boldsymbol{H}_{}$.}
\section{System Model}
\label{sec2}
\begin{figure}
\center
\includegraphics[totalheight=6.5cm,width=6.5cm]{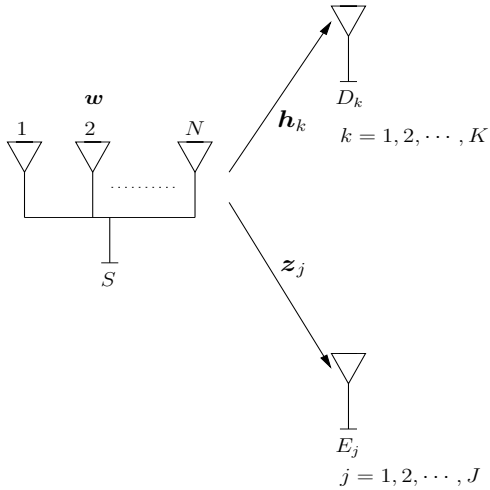}
\caption{System model for MISO wiretap channel with $K$ users and $J$ eavesdroppers.} 
\label{fig1}
\end{figure}

Consider a MISO wiretap channel as shown in Fig. \ref{fig1},
which consists 
of a source $S$ having $N$ transmit antennas, $K$ users $ \{ D_{1}, D_{2},\cdots,D_{K}\}$ each 
having single antenna, and $J$ {non-colluding} eavesdroppers $ \{ E_{1}, E_{2},\cdots,E_{J}\}$ each 
having single antenna. The complex channel gain vector
from $S$ to $D_{k}$ is 
denoted by $\boldsymbol{h}_{k} \in \mathbb{C}^{1 \times N}$, $1\leq k \leq K$.
Likewise, the complex channel gain vector from $S$ to $E_{j}$ is denoted by 
$\boldsymbol{z}_{j} \in \mathbb{C}^{1 \times N}$, $1\leq j \leq J$.
We assume that the channels
between $S$ to $D_{k}$s and those between $S$ to $E_{j}$s fade slowly and independently with 
$\boldsymbol{h}_{k} \sim \mathcal{CN}(\boldsymbol{0}, \boldsymbol{H}_{k})$, and
$\boldsymbol{z}_{j} \sim \mathcal{CN}(\boldsymbol{0}, \boldsymbol{Z}_{j})$. These channel gains
are assumed to be unknown at $S$.
We assume that the source
$S$ operates under total power constraint $P_{T}$.
The communication between $S$ and $D_{k}$s
happens in $n$ channel uses. The source $S$ transmits secret message $W_{}$ which is 
equiprobable over $\{1,2,\cdots,2^{nR_{s}} \}$. 
For each $W_{}$ drawn equiprobably from the set 
$\{1,2,\cdots,2^{nR_{s}} \}$, the source, using a stochastic encoder,
maps $W_{}$ to a codeword 
$\{ x_{i} \}^{n}_{i = 1}$
of length $n$, where each $x_{i} \in \mathbb{C}^{}$, i.i.d. $\sim \mathcal{CN}(0,1)$, and 
$\mathbb{E}[ \lvert x_{i} \rvert^{2}] = 1$.
Each codeword $\{ x_{i} \}^{n}_{i = 1}$ belongs to a collection of $2^{nR_{D}}$ codewords {(i.e., wiretap code)}
where 
$R_{D} \geq R_{s}$.
The source applies the complex weight 
$\boldsymbol{w} = [w_{1}, w_{2},\cdots,w_{N}]^{T} \in \mathbb{C}^{N \times 1}$
and transmits the weighted symbol which is $\boldsymbol{w}x_{i}$ in the $i$th channel use, $1 \leq i \leq n$.
Since the source is power constrained, this implies that 
\begin{eqnarray}
\parallel \hspace{-1mm} \boldsymbol{w} \hspace{-1mm} \parallel^{2} \ \leq \ P_{T}. \label{eqn1}
\end{eqnarray}
In the following, 
we will use $x$ to denote the symbols in the codeword 
$\{ x_{i} \}^{n}_{i = 1}$.
{Since the channel is slow fading and the CSI is unknown at the source $S$,
we define the non-outage event for a given $(R_{D}, R_{s})$ pair of the wiretap code, denoted by $\mathcal{E}$, and
impose the probability constraint on $\mathcal{E}$ as follows:
\begin{eqnarray}
\mathcal{E} = \Big\{ R_{D_{k}} \ \geq \ R_{D}, \quad \forall k \ = \ 1,2,\cdots,K, \quad \text{and} \nonumber \\ 
 R_{D} - R_{s} \ \geq \ R_{E_{j}}, \quad \forall j \ = \ 1,2,\cdots,J \Big \}, \label{eqn45} \\
\text{Pr} (\mathcal{E}) \ \geq \ (1-\epsilon), \label{eqn2}
\end{eqnarray}
where $(1- \epsilon)$ is the non-outage probability threshold, and 
$R_{D_{k}}$ and $R_{E_{j}}$ are the link information rates between
$S$ to $D_{k}$ and $S$ to $E_{j}$, respectively.
In other words, 
when the source selects the target code rate and target secrecy rate pair
of the wiretap code as $(R_{D}, R_{s})$,
the above constraint implies that, with probability greater than or equal to $(1- \epsilon)$,
all $D_{k}$s will be able to successfully decode the transmitted message while all $E_{j}$s
will be ignorant about the transmitted message.
}
We also note that when the CSI on all the links are known at $S$, 
the achievability of the secrecy rate $R_{s}$
is shown in \cite{ir11}.
Let $y_{D_{k}}$ and $y_{E_{j}}$ denote the received signals at $D_{k}$ and $E_{j}$, respectively.
We have
\begin{eqnarray}
y_{D_{k}} \ &=& \ \boldsymbol{h}_{k} \boldsymbol{w}x + \eta_{D_{k}}, \quad \forall k \ = \ 1,2,\cdots,K, \label{eqn3} \\
y_{E_{j}} \ &=& \ \boldsymbol{z}_{j} \boldsymbol{w}x + \eta_{E_{j}}, \quad \forall j \ = \ 1,2,\cdots,J, \label{eqn4}
\end{eqnarray}
where $\eta$s are noise components, assumed to be i.i.d. $ \sim \mathcal{CN}(0,N_0)$.

\section{Transmitter Optimization under Secrecy Constraint}
\label{sec3}
Using (\ref{eqn3}) and (\ref{eqn4}), and for a given $\boldsymbol{h}_{k}$ and $\boldsymbol{z}_{j}$, the information rates at $D_{k}$ and $E_{j}$
are obtained, respectively, as follows:
\begin{eqnarray}
R_{D_{k}} \ = \ I(x; \ y_{D_{k}}) \ = \ \log_{2} \Big( 1 + \frac{ \lvert \boldsymbol{h}_{k} \boldsymbol{w} \rvert^{2} } {N_{0}} \Big), \label{eqn5} \\
R_{E_{j}} \ = \ I(x; \ y_{E_{j}}) \ = \ \log_{2} \Big( 1 + \frac{ \lvert \boldsymbol{z}_{j} \boldsymbol{w} \rvert^{2} } {N_{0}} \Big). \label{eqn6}
\end{eqnarray}
Further, subject to the constraints in (\ref{eqn1}) and (\ref{eqn2}) and using (\ref{eqn5}) and (\ref{eqn6}), the optimization
problem to minimize the transmit power is as follows:
\begin{eqnarray}
\min_{\boldsymbol{w}} \ \parallel \hspace{-1mm} \boldsymbol{w} \hspace{-1mm} \parallel^{2} \label{eqn7} 
\end{eqnarray}
\begin{eqnarray}
\text{s.t.} \quad \parallel \hspace{-1mm} \boldsymbol{w} \hspace{-1mm} \parallel^{2} \ \leq \ P_{T},  \label{eqn8} \\
\text{Pr} \Big \{ \log_{2} \Big( 1 + \frac{ \lvert \boldsymbol{h}_{k} \boldsymbol{w} \rvert^{2} } {N_{0}} \Big) \ \geq \ R_{D}, \quad  
\forall k \ = \ 1,2,\cdots,K, \nonumber \\ 
R_{D} - R_{s} \ \geq \ \log_{2} \Big( 1 + \frac{ \lvert \boldsymbol{z}_{j} \boldsymbol{w} \rvert^{2} } {N_{0}} \Big), \quad 
\forall j \ = \ 1,2,\cdots,J \Big \} \nonumber \\ 
\geq \ (1-\epsilon).  \label{eqn8}
\end{eqnarray}
Since $\boldsymbol{h}_{k}$s and $\boldsymbol{z}_{j}$s are independent, we rewrite the constraint (\ref{eqn8}) in the following
equivalent product form:
\begin{eqnarray}
\prod^{K}_{k = 1} \text{Pr} \Big \{ \lvert \boldsymbol{h}_{k} \boldsymbol{w} \rvert^{2} \ \geq \ \big(2^{R_{D}} - 1\big)N_{0} \Big\} \nonumber \\
\prod^{J}_{j = 1} \text{Pr} \Big \{ \lvert \boldsymbol{z}_{j} \boldsymbol{w} \rvert^{2} \ \leq \ \big(2^{\big(R_{D} - R_{s}\big)} - 1\big)N_{0} \Big\} \ \geq \  (1-\epsilon). \label{eqn9}
\end{eqnarray}
{ We note that solving the optimization problem (\ref{eqn7}) in its original form is hard.
So, in order to simplify the analysis, we replace the product probability 
constraint in (\ref{eqn9}) with 
the following $K + J$ individual probability constraints:}
\begin{eqnarray}
\forall k \ = \ 1,2,\cdots,K, \quad \text{and} \quad \forall j \ = \ 1,2,\cdots,J, \nonumber \\
\text{Pr} \Big \{ \lvert \boldsymbol{h}_{k} \boldsymbol{w} \rvert^{2} \ \geq \ \big(2^{R_{D}} - 1\big)N_{0} \Big\} \ \geq \ (1-\epsilon)^{ \frac{1}{K + J} }, \label{eqn10} \\
\text{Pr} \Big \{ \lvert \boldsymbol{z}_{j} \boldsymbol{w} \rvert^{2} \ \leq \ \big(2^{\big(R_{D} - R_{s}\big)} - 1\big)N_{0} \Big\} \ \geq \ (1-\epsilon)^{ \frac{1}{K + J} }. \label{eqn11}
\end{eqnarray}
{ We also note that any $\boldsymbol{w}$ which satisfies all $K+J$ constraints in (\ref{eqn10}) and (\ref{eqn11})
will also satisfy the product probability constraint in (\ref{eqn9}). However, the converse may not always be true.}
Further, since $\boldsymbol{h}_{k} \boldsymbol{w}$ in (\ref{eqn10}) and $\boldsymbol{z}_{j} \boldsymbol{w}$ in (\ref{eqn11}) are { linear
transformations of circularly symmetric complex Gaussian random vectors},
$\boldsymbol{h}_{k} \boldsymbol{w}$ and $\boldsymbol{z}_{j} \boldsymbol{w}$ are also circularly symmetric complex Gaussian random variables,
i.e., $\boldsymbol{h}_{k} \boldsymbol{w} \sim \mathcal{CN} \big(0, \ \boldsymbol{w}^{\ast}\boldsymbol{H}_{k}\boldsymbol{w} \big)$, and
$\boldsymbol{z}_{j} \boldsymbol{w} \sim \mathcal{CN} \big(0, \ \boldsymbol{w}^{\ast}\boldsymbol{Z}_{j}\boldsymbol{w} \big)$.
This further implies that $\lvert \boldsymbol{h}_{k} \boldsymbol{w} \rvert^{2}$ and $\lvert \boldsymbol{z}_{j} \boldsymbol{w} \rvert^{2}$
are exponential random variables, i.e.,
\begin{eqnarray}
\lvert \boldsymbol{h}_{k} \boldsymbol{w} \rvert^{2} \sim \frac{1}{\boldsymbol{w}^{\ast}\boldsymbol{H}_{k}\boldsymbol{w}} \exp^{- \frac{\lambda}{\boldsymbol{w}^{\ast}\boldsymbol{H}_{k}\boldsymbol{w}}}, \quad \lambda \ \geq \ 0,   \label{eqn12}\\
\lvert \boldsymbol{z}_{j} \boldsymbol{w} \rvert^{2} \sim \frac{1}{\boldsymbol{w}^{\ast}\boldsymbol{Z}_{j}\boldsymbol{w}} \exp^{- \frac{\lambda}{\boldsymbol{w}^{\ast}\boldsymbol{Z}_{j}\boldsymbol{w}}}, \quad \lambda \ \geq \ 0. \label{eqn13} 
\end{eqnarray}
{Using (\ref{eqn12}) and (\ref{eqn13}), and by following standard integration steps, 
we get the following equivalent simplified inequalities for the probability constraints in (\ref{eqn10}) and (\ref{eqn11}):}
\begin{eqnarray}
\forall k \ = \ 1,2,\cdots,K, \quad \boldsymbol{w}^{\ast}\boldsymbol{H}_{k}\boldsymbol{w} \ \geq \  a, \label{eqn14} \\
\forall j \ = \ 1,2,\cdots,J, \quad \boldsymbol{w}^{\ast}\boldsymbol{Z}_{j}\boldsymbol{w} \ \leq \  b, \label{eqn15}
\end{eqnarray}
where $a = \frac{(2^{R_{D}} - 1)N_{0}}{-\ln (1 - \epsilon)^{\frac{1}{(K+J)}}}$ and 
$b = \frac{(2^{(R_{D} - R_{s})} - 1)N_{0}}{-\ln (1 - (1-\epsilon)^{\frac{1}{(K+J)}} )}$.
Replacing the constraint in (\ref{eqn9}) with (\ref{eqn14}) and (\ref{eqn15}), we get
the following { upper bound} for the optimization problem (\ref{eqn7}):
\begin{eqnarray}
\min_{\boldsymbol{w}} \ \parallel \hspace{-1mm} \boldsymbol{w} \hspace{-1mm} \parallel^{2} \label{eqn16} 
\end{eqnarray}
\begin{eqnarray}
\text{s.t.} \quad \parallel \hspace{-1mm} \boldsymbol{w} \hspace{-1mm} \parallel^{2} \ \leq \ P_{T},  \label{eqn17} \\
\forall k \ = \ 1,2,\cdots,K, \quad \boldsymbol{w}^{\ast}\boldsymbol{H}_{k}\boldsymbol{w} \ \geq \  a, \label{eqn18} \\
\forall j \ = \ 1,2,\cdots,J, \quad \boldsymbol{w}^{\ast}\boldsymbol{Z}_{j}\boldsymbol{w} \ \leq \  b, \label{eqn19}
\end{eqnarray}
We solve the above problem for the following two cases.
\subsection{All $\boldsymbol{H}_{k}$s and $\boldsymbol{Z}_{j}$s are diagonal matrices}
\label{sec3A}
When all $\boldsymbol{H}_{k}$s and $\boldsymbol{Z}_{j}$s are diagonal positive semidefinite matrices,
the optimization problem (\ref{eqn16}) can be written as the following equivalent
linear optimization problem:
\begin{eqnarray}
\min_{P_{1}, P_{2}, \cdots, P_{N}} \ \sum^{N}_{m = 1} P_{m}\label{eqn34} 
\end{eqnarray}
\begin{eqnarray}
\text{s.t.} \quad \forall m \ = \ 1,2,\cdots,N, \quad P_{m} \geq 0, \quad \sum^{N}_{m = 1} P_{m} \ \leq \ P_{T},  \label{eqn35} \\
\forall k \ = \ 1,2,\cdots,K, \quad \sum^{N}_{m = 1} P_{m}{H}^{mm}_{k} \ \geq \ a, \label{eqn35} \\
\forall j \ = \ 1,2,\cdots,J, \quad \sum^{N}_{m = 1} P_{m}{Z}^{mm}_{j} \ \leq \ b, \label{eqn36}
\end{eqnarray}
where 
$P_{m} = |w_{m}|^{2}$, 
$\boldsymbol{H}_{k} = \diagonal{([{H}^{11}_{k}, {H}^{22}_{k},\cdots,{H}^{NN}_{k}]^{T})} \succeq \boldsymbol{0}$, and
$\boldsymbol{Z}_{j} = \diagonal{([{Z}^{11}_{j}, {Z}^{22}_{j},\cdots,{Z}^{NN}_{j}]^{T})} \succeq \boldsymbol{0}$.
The above problem can be easily solved using linear optimization techniques.
Having obtained $P_{1}, P_{2}, \cdots, P_{N}$, the beamforming vector $\boldsymbol{w}$ is $[\sqrt{P_{1}},\sqrt{P_{2}},\cdots,\sqrt{P_{N}}]^{T}$.
\subsection{Some of $\boldsymbol{H}_{k}$s or $\boldsymbol{Z}_{j}$s are not diagonal matrices}
\label{sec3B}
Here, we consider the general case when $\boldsymbol{H}_{k}$s and $\boldsymbol{Z}_{j}$s are Hermitian 
positive semidefinite matrices and some of $\boldsymbol{H}_{k}$s or $\boldsymbol{Z}_{j}$s
are not diagonal. Define $\boldsymbol{W} \Define \boldsymbol{w}\boldsymbol{w}^{\ast}$. 
We rewrite the optimization problem (\ref{eqn16})
into the following equivalent form:
\begin{eqnarray}
\min_{\boldsymbol{W}} \ \Tr{(\boldsymbol{W})} \label{eqn20} 
\end{eqnarray}
\begin{eqnarray}
\text{s.t.} \quad \boldsymbol{W} \succeq \boldsymbol{0}, \quad rank(\boldsymbol{W}) = 1, \quad 
\Tr{(\boldsymbol{W})} \ \leq \ P_{T},  \label{eqn21} \\
\forall k \ = \ 1,2,\cdots,K, \quad \Tr{(\boldsymbol{W}\boldsymbol{H}_{k})} \ \geq \ a, \label{eqn22} \\
\forall j \ = \ 1,2,\cdots,J, \quad \Tr{(\boldsymbol{W}\boldsymbol{Z}_{j})} \ \leq \ b. \label{eqn23}
\end{eqnarray}
The above optimization problem is a non-convex optimization problem.
However, by relaxing the $rank(\boldsymbol{W}) = 1$ constraint, the above problem can be solved using 
semidefinite programming techniques \cite{ir15}. But the solution $\boldsymbol{W}$ of 
the above rank relaxed optimization problem may not have rank 1.
This can be easily seen from the KKT conditions of the rank relaxed optimization problem
which we discuss in the Appendix. 
We now take the rank-1 approximation as follows. Let $\boldsymbol{w}_{0}$ be the unit-norm 
eigen direction corresponding to the largest eigen value of $\boldsymbol{W}$.
We substitute $\boldsymbol{W} = P \boldsymbol{w}_{0} \boldsymbol{w}^{\ast}_{0}$ in the above
rank relaxed optimization problem and solve the resulting linear optimization problem for unknown $P$, 
i.e.,
\begin{eqnarray}
\min_{P} \ P \label{eqn24} 
\end{eqnarray}
\begin{eqnarray}
\text{s.t.} \quad  0 \ \leq \ P \leq \ P_{T},  \label{eqn25} \\
\forall k \ = \ 1,2,\cdots,K, \quad P\boldsymbol{w}^{\ast}_{0}\boldsymbol{H}_{k}\boldsymbol{w}^{}_{0} \ \geq \ a, \label{eqn26} \\
\forall j \ = \ 1,2,\cdots,J, \quad P\boldsymbol{w}^{\ast}_{0}\boldsymbol{Z}_{j}\boldsymbol{w}^{}_{0} \ \leq \ b. \label{eqn27}
\end{eqnarray}
Having obtained the transmit power $P$ from (\ref{eqn24}), 
the beamforming vector is $\sqrt{P} \boldsymbol{w}_{0}$. \\

{\em Remark 1:}
We note that when the channel CSI $\boldsymbol{h}_{k}$ on all $D_{k}$s are perfectly known at the source $S$,
the constraints (\ref{eqn22}) and (\ref{eqn23}) in the optimization problem (\ref{eqn20}) should be replaced
with the following constraints, respectively:
\begin{eqnarray}
\Tr{(\boldsymbol{W}^{} \boldsymbol{h}^{\ast}_{k} \boldsymbol{h}^{}_{k} )} \ \geq \ (2^{R_{D}} - 1)N_{0}, \label{eqn37} \\
\Tr{(\boldsymbol{W}\boldsymbol{Z}_{j})} \ \leq \ \frac{(2^{(R_{D} - R_{s})} - 1)N_{0}}{-\ln (1 - (1-\epsilon)^{\frac{1}{J}} )}. \label{eqn38} 
\end{eqnarray}

{\em Remark 2:}
When the source transmits the symbol $x$ from an equiprobable complex finite alphabet set 
${\mathbb{A}}=\{a_1, a_2,\cdots,a_M \}$ of size $M$ (e.g., $M$-ary) with ${\mathbb{E}}[x] = 0$ and ${\mathbb{E}}[|x|^{2}] = 1$,
the information rates in 
(\ref{eqn5}) and 
(\ref{eqn6}) 
can be written in the following forms, respectively:
\begin{eqnarray}
R_{D_{k}} \ = \ I(x; \ y_{D_{k}}) \ = \ I\bigg( \frac{ \lvert \boldsymbol{h}_{k} \boldsymbol{w} \rvert^{2} } {N_{0}} \bigg), \label{eqn_miso_slowfading_43} \\
R_{E_{j}} \ = \ I(x; \ y_{E_{j}}) \ = \ I\bigg( \frac{ \lvert \boldsymbol{z}_{j} \boldsymbol{w} \rvert^{2} } {N_{0}} \bigg), \label{eqn_miso_slowfading_44}
\end{eqnarray}
where
\begin{eqnarray}
I(\rho) \ \ \Define \ \ \frac{1}{M} \sum\limits^{M}_{l = 1} \int p_{n}\big(y_{_{}} - \sqrt{\rho} a_{l}\big) \nonumber \\
\log_{2} \frac{p_{n}(y_{_{}} - \sqrt{\rho} a_{l})}{\frac{1}{M} \sum\limits^{M}_{m = 1}p_{n}(y_{_{}} - \sqrt{\rho} a_{m})} d y_{_{}}, \label{eqn_miso_slowfading_45}
\end{eqnarray}
and $p_n(\theta) = \frac{1}{\pi} e^{{{-\mid \theta \mid}^{2}}}$. 
Using the fact that the mutual information function, $I(\rho)$, is a strictly-increasing concave function in $\rho$ \cite{ref_52, ref_53},
$K+J$ constraints in 
(\ref{eqn18}) and
(\ref{eqn19}) 
can be written in the following forms, respectively:
\begin{eqnarray}
\forall k \ = \ 1,2,\cdots,K, \quad \boldsymbol{w}^{\ast}\boldsymbol{H}_{k}\boldsymbol{w} \ \geq \  a, \label{eqn_miso_slowfading_46} \\
\forall j \ = \ 1,2,\cdots,J, \quad \boldsymbol{w}^{\ast}\boldsymbol{Z}_{j}\boldsymbol{w} \ \leq \  b_{}, \label{eqn_miso_slowfading_47}
\end{eqnarray}
where $a = \frac{I^{-1}(R_{D})N_{0}}{-\ln (1 - \epsilon)^{\frac{1}{(K+J)}}}$ and 
$b_{} = \frac{I^{-1}(R_{D} - R_{s})N_{0}}{-\ln (1 - (1-\epsilon)^{\frac{1}{(K+J)}} )}$.
With finite alphabet input, the optimization problem (\ref{eqn16})
should be solved subject to the constraints in 
(\ref{eqn_miso_slowfading_46}) and
(\ref{eqn_miso_slowfading_47}).
\section{Results and Discussions}
\label{sec4}
{We have evaluated the secrecy rate through simulation with the following system parameters:
$N = 3, \ K = 2, \ J = 1,2,3, \ N_{0} = 1, \ \epsilon = 0.1$, and $P_{T} = 12$ dB.
We consider the scenarios discussed in Section \ref{sec3A} and Section \ref{sec3B}.

{\em Scenario of Section \ref{sec3B}:}
We have used the following positive definite channel covariance matrices in the simulations:

{\footnotesize
\begin{eqnarray}
\boldsymbol{H}_1 =
\left[\footnotesize
\begin{array}{cc}
2.1670, \ \  0.1806 + 0.0183i,  \ \ -0.1453 - 0.3101i \\   0.1806 - 0.0183i, \ \ 1.9165, \ \ 0.0696 + 0.3374i \\ -0.1453 + 0.3101i, \ \ 0.0696 - 0.3374i, \ \ 1.4180
\end{array}  \right] \succ \boldsymbol{0} \label{eqn40} \\
\boldsymbol{H}_2 =
\left[\footnotesize
\begin{array}{cc}
1.9834, \ \ -0.2001 + 0.0250i, \ \ 0.0470 - 0.3424i \\    -0.2001 - 0.0250i, \ \ 1.3867, \ \ 0.0149 - 0.2083i \\  0.0470 + 0.3424i, \ \  0.0149 + 0.2083i, \ \ 1.4323
\end{array}  \right] \succ \boldsymbol{0} \label{eqn41} \\
\boldsymbol{Z}_1 =
\left[\footnotesize
\begin{array}{cc}
0.0043, \ \ 0.0010 - 0.0003i, \ \ 0.0013 + 0.0009i \\      0.0010 + 0.0003i, \ \ 0.0074, \ \ -0.0011 - 0.0029i \\  0.0013 - 0.0009i, \ \ -0.0011 + 0.0029i, \ \ 0.0079
\end{array}  \right] \succ \boldsymbol{0} \label{eqn42} \\
\boldsymbol{Z}_2 =
\left[\footnotesize
\begin{array}{cc}
0.0069, \ \ 0.0004 - 0.0029i, \ \ -0.0014 + 0.0014i \\     0.0004 + 0.0029i, \ \ 0.0070, \ \ -0.0019 - 0.0002i \\ -0.0014 - 0.0014i, \ \ -0.0019 + 0.0002i, \ \ 0.0086
\end{array}  \right] \succ \boldsymbol{0} \label{eqn43} \\
\boldsymbol{Z}_3 =
\left[\footnotesize
\begin{array}{cc}
0.0090, \ \ -0.0026 + 0.0006i, \ \ 0.0011 - 0.0009i \\   -0.0026 - 0.0006i, \ \ 0.0064, \ \ -0.0013 + 0.0018i \\  0.0011 + 0.0009i, \ \ -0.0013 - 0.0018i, \ \ 0.0054
\end{array}  \right] \succ \boldsymbol{0} \label{eqn44}
\end{eqnarray}
}

\hspace{-5.0mm} For a given $(R_{D}, R_{s})$ pair, 
we solve the semidefinite rank relaxed optimization problem (\ref{eqn20}) using the 
tools in \cite{ir16, ir17}. We numerically observe that, for any feasible $(R_{D}, R_{s})$ pair,
the solution $\boldsymbol{W}$ 
of the rank relaxed optimization problem (\ref{eqn20}) has rank 1.
This implies that for such channel realizations, rank-1 approximation is not needed. 
In Fig \ref{fig2}(a),
\begin{figure}[htb]
\begin{minipage}[b]{.48\linewidth}
\centering
\centerline{\epsfig{figure=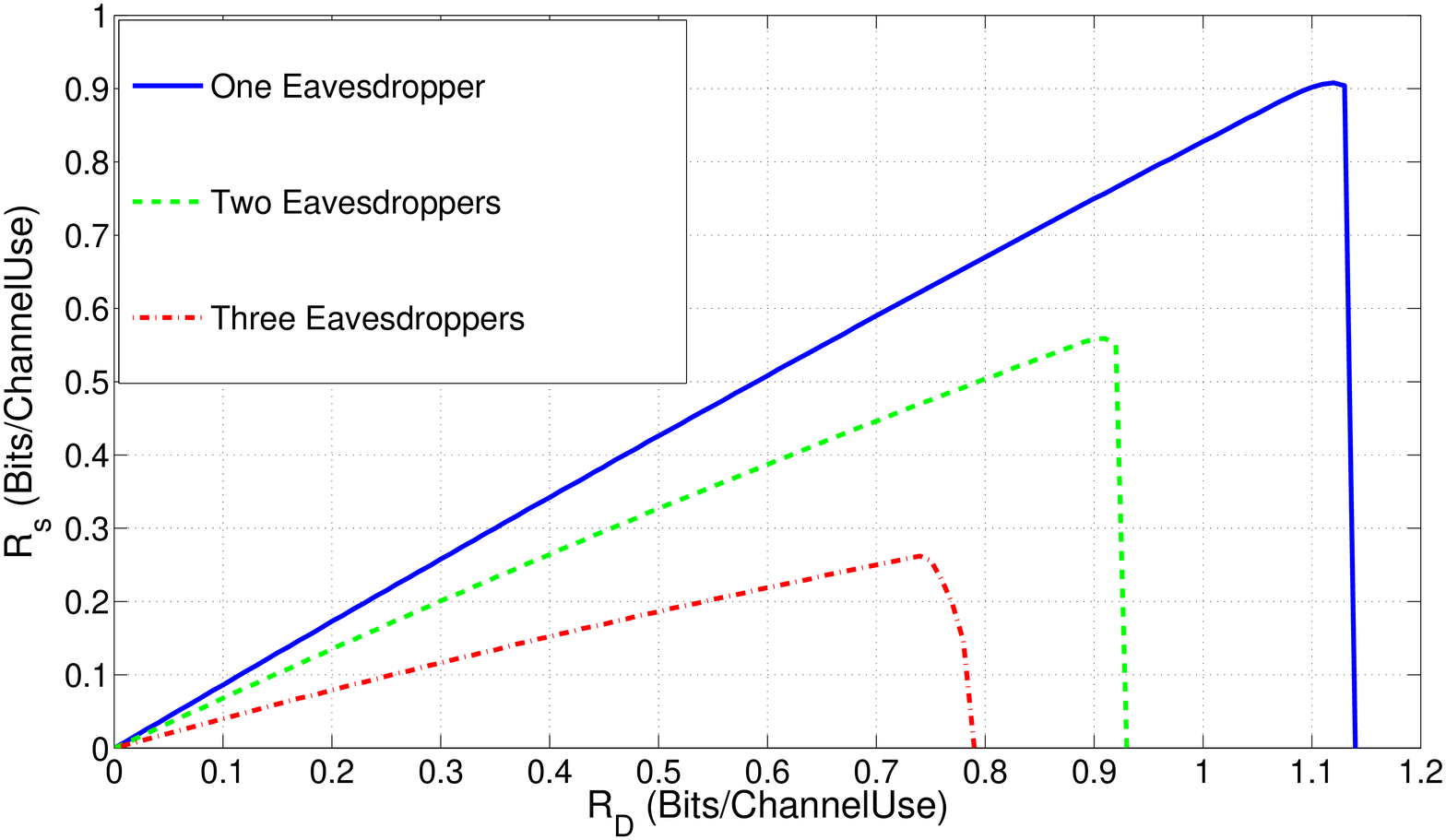,width=4.65cm,height=8.00cm}}
\centerline{(a) $R_{s}$ vs $R_{D}$.} \medskip
\end{minipage}
\hfill
\begin{minipage}[b]{0.48\linewidth}
\centering
\centerline{\epsfig{figure=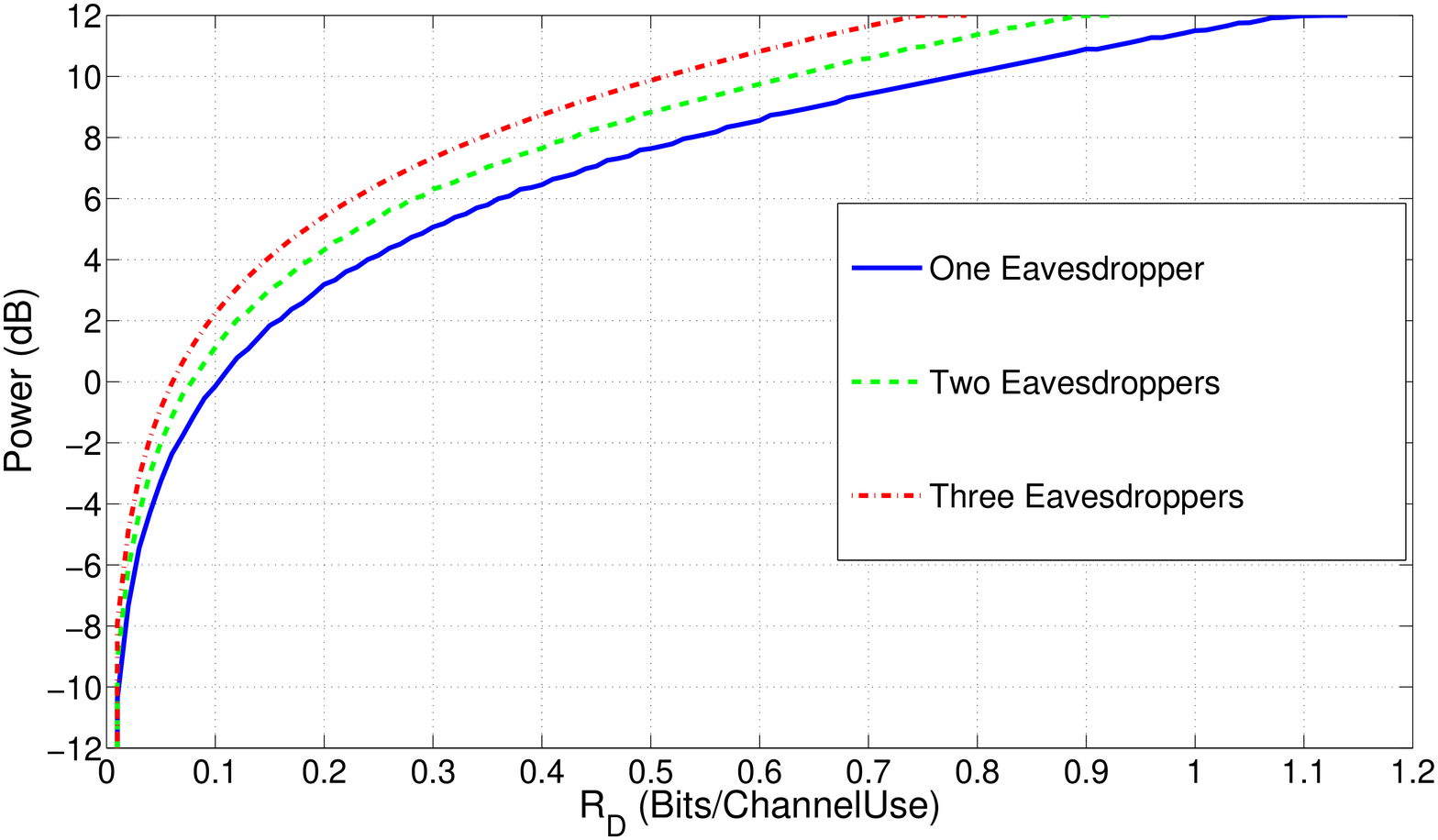,width=4.65cm, height=8.00cm}}
\centerline{(b) Transmit Power vs $R_{D}$.}\medskip
\end{minipage}
\caption{$R_{s}$ vs $R_{D}$ and Transmit Power vs $R_{D}$ in MISO wiretap channel with $N = 3, \ K = 2, \ J = 1,2,3, \ N_{0} = 1, \ \epsilon = 0.1$ and $P_{T} = 12$ dB,
and non-diagonal covariance matrices.}
\label{fig2}
\end{figure}
we plot the maximum achievable $R_{s}$ vs $R_{D}$. In Fig \ref{fig2}(b), 
we plot the corresponding mimimum transmit power vs $R_{D}$.
We observe that the maximum achievable secrecy rate $R_{s}$ and 
the corresponding minimum transmit power increases with increase in $R_{D}$.
The secrecy rate drops to zero when the entire available power, $P_{T} = 12$ dB,
is used.

{\em Scenario of Section \ref{sec3A}:} Here, we take
$\boldsymbol{H}_{1}$, 
$\boldsymbol{H}_{2}$,
$\boldsymbol{Z}_{1}$,
$\boldsymbol{Z}_{2}$, and 
$\boldsymbol{Z}_{3}$ 
as the diagonal approximation of covariance matrices in
(\ref{eqn40}), 
(\ref{eqn41}), 
(\ref{eqn42}), 
(\ref{eqn43}), and
(\ref{eqn44}), respectively.
We solve the linear optimization problem (\ref{eqn34}) using the tools in \cite{ir16, ir17}, and
we plot the maximum achievable $R_{s}$ vs $R_{D}$ and
the corresponding mimimum transmit power vs $R_{D}$
in Fig \ref{fig3}(a)
\begin{figure}[htb]
\begin{minipage}[b]{.48\linewidth}
\centering
\centerline{\epsfig{figure=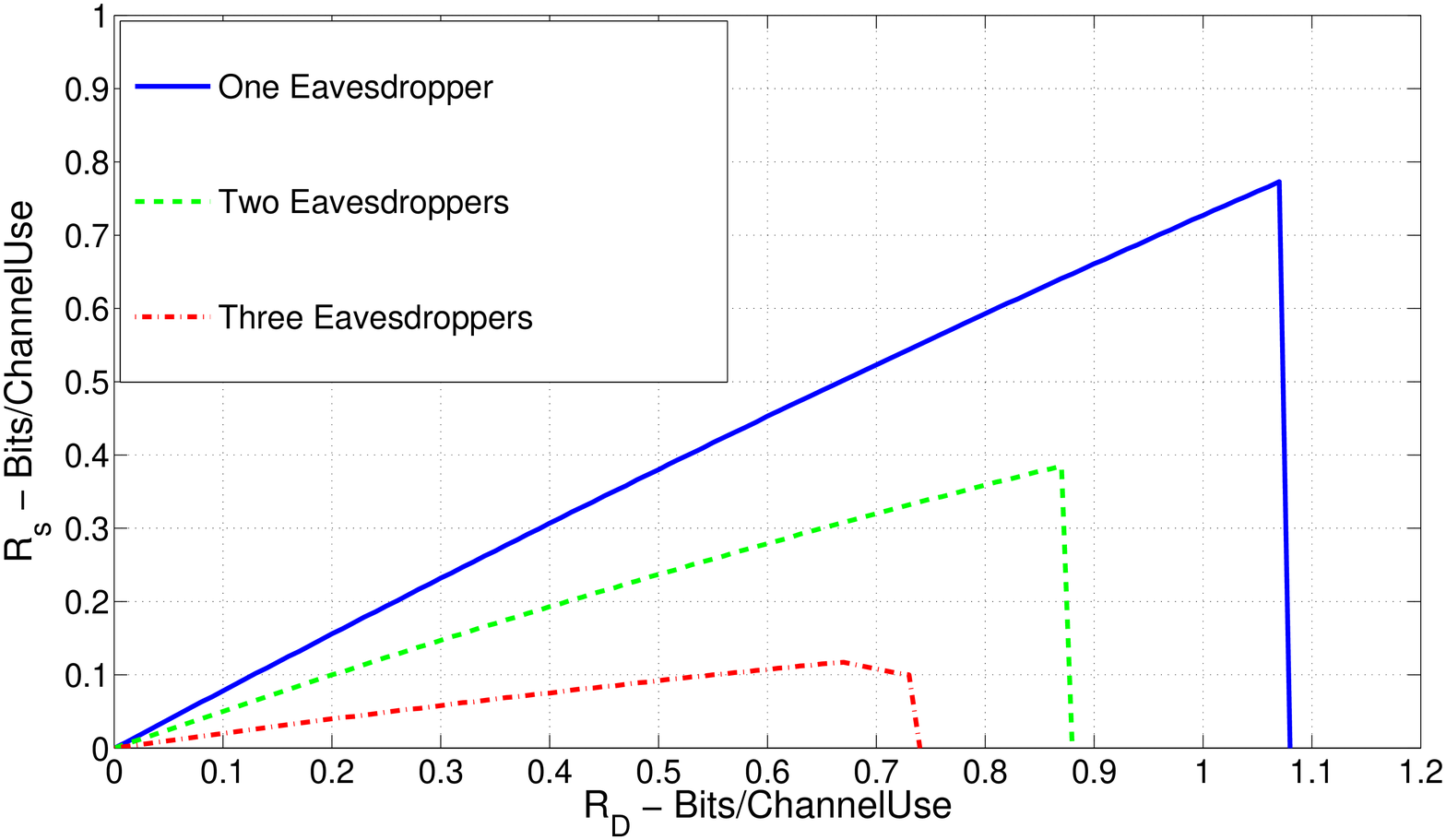,width=4.65cm,height=8.00cm}}
\centerline{(a) $R_{s}$ vs $R_{D}$.} \medskip
\end{minipage}
\hfill
\begin{minipage}[b]{0.48\linewidth}
\centering
\centerline{\epsfig{figure=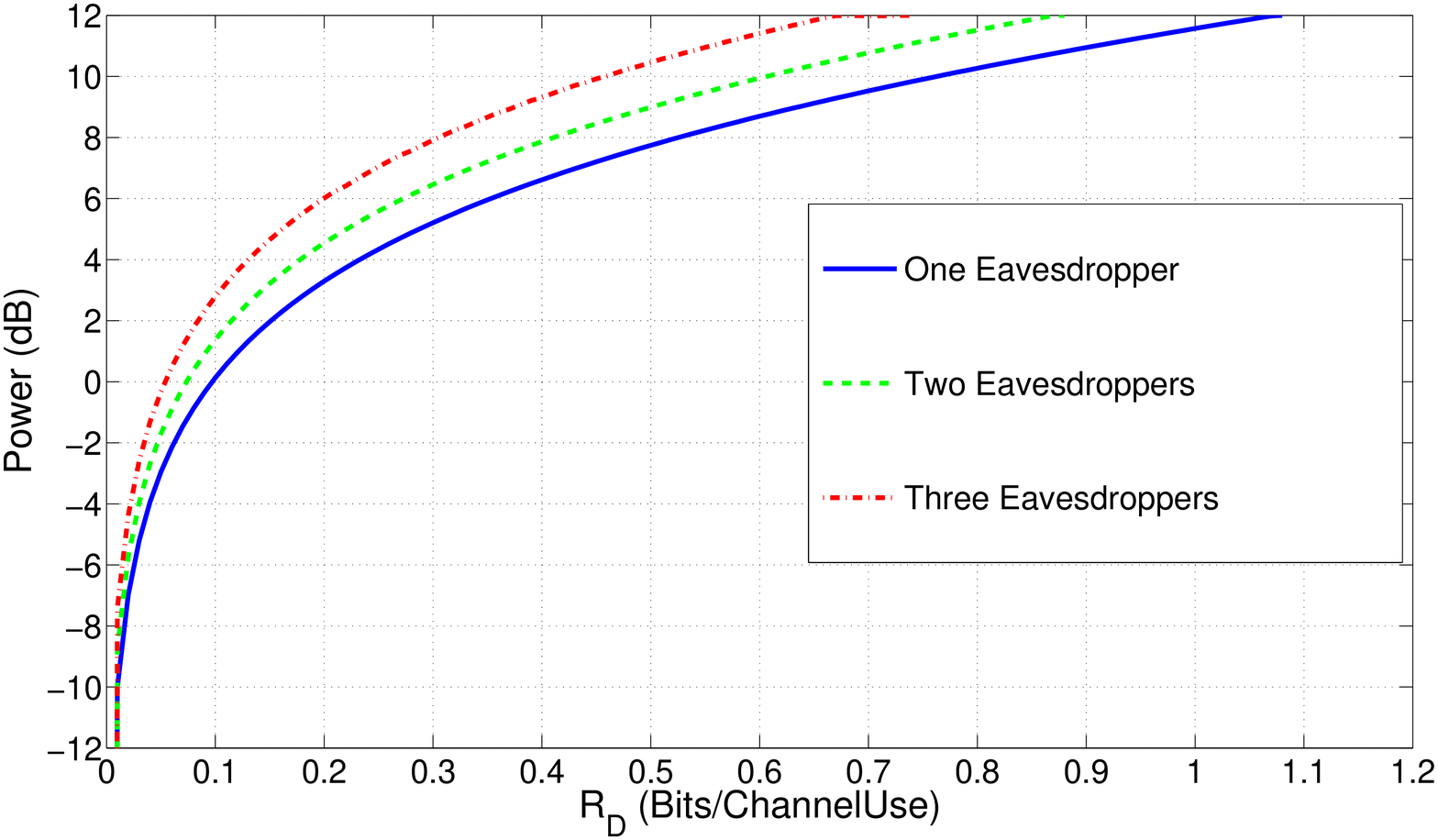,width=4.65cm, height=8.00cm}}
\centerline{(b) Transmit Power vs $R_{D}$.}\medskip
\end{minipage}
\caption{$R_{s}$ vs $R_{D}$ and Transmit Power vs $R_{D}$ in MISO wiretap channel with $N = 3, \ K = 2, \ J = 1,2,3, \ N_{0} = 1, \ \epsilon = 0.1$ and $P_{T} = 12$ dB,
and diagonal covariance matrices.} 
\label{fig3}
\end{figure}
and Fig \ref{fig3}(b), respectively.
As in Fig \ref{fig2}(a) and Fig \ref{fig2}(b),
we observe that the maximum achievable secrecy rate $R_{s}$ and 
the corresponding minimum transmit power increases with increase in $R_{D}$.
The secrecy rate drops to zero when the entire available power, $P_{T} = 12$ dB,
is used.
}
\section{Conclusions}
\label{sec5}
We considered the transmitter optimization problem
in slow fading MISO wiretap channel. 
Secret message transmitted by the source was intended for $K$ users in the presence of $J$ eavesdroppers.
For a given code rate and secrecy rate pair of the wiretap code, 
denoted by $(R_{D}, R_{s})$,
we defined the non-outage event and
minimized the transmit power subject to the total power constraint
and satisfying the probability of the non-outage event
to be greater than a desired threshold $(1-\epsilon)$.
We obtained the achievable ($R_{D}, R_{s}$) region
and the transmit beamforming vector.
\section*{Appendix}
\label{sec6}
In this appendix, we analyze the rank of the optimal solution $\boldsymbol{W}$ of the rank relaxed optimization
problem (\ref{eqn20}). We take the Lagrangian \cite{ir15} of the rank relaxed optimization problem
(\ref{eqn20}) as follows:
\begin{eqnarray}
\ell \big( \boldsymbol{W}, \ \boldsymbol{\Lambda},  \ \lambda, \ \mu_{k}, \ \nu_{j} \big) \ = \ \Tr{(\boldsymbol{W})} 
- \Tr{(\boldsymbol{\Lambda}\boldsymbol{W})} \nonumber \\
+ \ \lambda \big( \Tr{(\boldsymbol{W})} - P_{T} \big) 
+ \sum^{K}_{k = 1} \mu_{k} \big( a - \Tr{(\boldsymbol{W} \boldsymbol{H}_{k})} \big) \nonumber \\ 
+ \sum^{J}_{j = 1} \nu_{j} \big( \Tr{(\boldsymbol{W} \boldsymbol{Z}_{j})} - b  \big), \label{eqn28}
\end{eqnarray}
where $\boldsymbol{\Lambda} \succeq \boldsymbol{0}$, $\lambda \geq 0$, $\mu_{k} \geq 0$, and $\nu_{j} \geq 0$
are Lagrangian multipliers. The KKT conditions are as follows:
\begin{itemize}

\vspace{2mm}
\item[K1.] 
All the constraints in (\ref{eqn21}), (\ref{eqn22}), and  (\ref{eqn23}) excluding the constraint $rank(\boldsymbol{W}) = 1$,

\vspace{2mm}
\item[K2.] 
$\Tr{(\boldsymbol{\Lambda}\boldsymbol{W})} = 0$. Since $\boldsymbol{\Lambda} \succeq \boldsymbol{0}$ and $\boldsymbol{W} \succeq \boldsymbol{0}$,
this implies that $\boldsymbol{\Lambda}\boldsymbol{W} = \boldsymbol{0}$,

\vspace{2mm}
\item[K3.]  
$\lambda \big( \Tr{(\boldsymbol{W}) - P_{T} \big)} = 0$, 

\vspace{2mm}
\item[K4.] 
$\forall k = 1,2,\cdots,K, \quad$ $\mu_{k} \big( a - \Tr{(\boldsymbol{W} \boldsymbol{H}_{k})} \big) = 0$,

\vspace{2mm}
\item[K5.] 
$\forall j = 1,2,\cdots,J, \quad$ $\nu_{j} \big( \Tr{(\boldsymbol{W} \boldsymbol{Z}_{j})} - b \big) = 0$,

\vspace{2mm}
\item[K6.] 
$\frac{\partial \ell}{\partial \boldsymbol{W}} = \boldsymbol{0}$ implies that $\boldsymbol{\Lambda} = (1 + \lambda)\boldsymbol{I} - \sum^{K}_{k = 1} \mu_{k} \boldsymbol{H}_{k} + \sum^{J}_{j = 1} \nu_{j} \boldsymbol{Z}_{j} \succeq \boldsymbol{0}$,

\end{itemize}

The KKT conditions (K2), (K6), (K4), and (K5) imply that $(1 + \lambda) \Tr{(\boldsymbol{W})} - \sum^{K}_{k = 1} \mu_{k} a + \sum^{J}_{j = 1} \nu_{j} b  = 0$. For $\boldsymbol{W} \neq \boldsymbol{0}$,
this further implies that not all $\mu_{k}$s can be zero simultaneously. With this, we rewrite (K6) in the following form:
\begin{eqnarray}
\boldsymbol{\Lambda} + \sum^{K}_{k = 1} \mu_{k} \boldsymbol{H}_{k}  \ = \ (1 + \lambda)\boldsymbol{I} + \sum^{J}_{j = 1} \nu_{j} \boldsymbol{Z}_{j} \ \succ \ \boldsymbol{0}. \label{eqn29}
\end{eqnarray}
The above equation implies that $rank( \boldsymbol{\Lambda} + \sum^{K}_{k = 1} \mu_{k} \boldsymbol{H}_{k} ) = N$. 
This further implies that $rank( \boldsymbol{\Lambda} ) \geq N - rank(\sum^{K}_{k = 1} \mu_{k} \boldsymbol{H}_{k} )$.
(K2) implies that $rank( \boldsymbol{W} ) \leq rank(\sum^{K}_{k = 1} \mu_{k} \boldsymbol{H}_{k} )$
(assuming $\boldsymbol{W} \neq \boldsymbol{0}$). This means that the rank of $\boldsymbol{W}$ may not be one.

For the special case when $K = 1$ and $\boldsymbol{H}_{1}$ is a rank one positive semidefinite matrix,
(\ref{eqn29}) implies that $rank( \boldsymbol{\Lambda} ) \geq N-1$. 
Assuming $\boldsymbol{W} \neq \boldsymbol{0}$, (K2) further implies that  $rank( \boldsymbol{\Lambda} ) = N-1$,
and $rank( \boldsymbol{W} ) = 1$.

\end{document}